\documentclass[a4paper]{PoS}

\usepackage{lineno}

\title{Detection of Near Horizontal Muons with the HAWC Observatory}

\ShortTitle{Detection of Near Horizontal Muons}

\author{Ahron S. Barber$^{a}$, David B. Kieda$^{a}$ 
and \speaker{R. Wayne Springer}$^{a}$ for the HAWC Collaboration$^{b}$ \\
        \llap{$^{a}$}Department of Physics and Astronomy, 
        University of Utah, Salt Lake City, UT, USA \\
        \llap{$^{b}$}For a complete author list, see 
        www.hawc-observatory.org/collaboration/icrc2017.php\\
        Email: \email{ahron.barber@utah.edu}, \email{dave.kieda@utah.edu}, \email{wayne.springer@utah.edu}}

\abstract{The HAWC (High Altitude Water Cherenkov) gamma ray observatory is able to observe muons with nearly horizontal trajectories. HAWC is located at an altitude of 4100 meters a.s.l. on the Sierra Negra volcano in Mexico. The HAWC detector is composed of 300 water tanks, each 7.3 m in diameter and 4.5 m tall, densely packed over a physical area of 22,000 m$^{2}$. Previous and current experiments have observed high zenith angle (near horizontal) muons at or near sea level. HAWC operates as a hodoscope able to observe multi-TeV muons at zenith angles greater than 75 degrees. This is the first experiment to measure near horizontal muons at high altitude and with large ($\geq$ 10 m) separations for multiple muons. These muons are distinguishable from extensive air showers by observing near horizontal particles propagating with the speed of light. The proximity of Sierra Negra and Pico de Orizaba volcanoes provides an additional measurement of muons with rock overburdens of several km water equivalent. We will present the angular distribution and rate at which HAWC observes these muon events.}

\FullConference{35th International Cosmic Ray Conference, ICRC2017\\
		10-20 July, 2017\\
		Bexco, Busan, Korea}

\begin{document}

\section{INTRODUCTION}

HAWC is the High Altitude Water Cherenkov Observatory, which is designed to detect VHE Gamma-Ray and Cosmic-Ray extensive air shower events. These air shower events occur within HAWC at a rate of 25 kHz, allowing it to pursue its primary science objectives. The objectives are to study the Milky Way galactic plane gamma ray sources, extra-galactic sources such as Markarian 421, and the observed large and small-scale cosmic ray anisotropies. These standard HAWC analyses consider events with measured zenith angles are typically less than 60 degrees. In this study, we investigate events that contain muons with trajectories that are nearly horizontal at zenith angles greater than 75 degrees.

Studying the muons, which reach the Earth's surface, has a history stretching back to their discovery in the field of cosmic rays. The experiments range from simple cloud chambers, to spectrometers like MUTRON \cite{MUTRON}, to detectors like MACRO \cite{MACRO} placed in old mines. These past experiments look for single and multiple muon events. In recent history there has been a resurgence in looking at these muons to investigate what contribution they may have for the cosmic ray community. One of the primary science goals for these muon experiments is to provide possible ground based observations for cosmic ray composition studies above $10^{15}$ electronvolts (PeV). Another interesting science question is the muon excess in Extensive Air Shower (EAS) most recently observed by the Pierre Auger Observatory \cite{PAuger1}, the models used to generate the EAS underestimate the hadronic component for the shower when pushed past what is tested in the lab. 

Studying nearly horizontal muons started in the mid 1970's \cite{MUTRON} and continue through to today. The most recent experiment, similar to our study is the Decor-Nevod experiment, which has been studying near horizontal muon bundles, containing more than 3 muons and up to around 80 muons. These bundles of muons are believed to originate from cosmic ray primaries with energies above 100 PeV \cite{DecorNevod}. The Decor-Nevod detector is located nearly at sea level (Moscow,Russia) with dimensions 9 x 9 x 26 $m^{3}$.  The muon bundles that they observe are highly compact and their detector segmentation has been optimized to detect and resolve the constituent muons. The greater lateral extent, 150 m, of the HAWC detector provides greater sensitive area. High muon multiplicities in HAWC may be distinguishable via charge measurements in addition to spatially resolving multiple tracks.

Muon tomography for imaging thick objects such as mountains and volcanoes has recently become an interesting subject. Volcanoes are particularly interesting as the technique is being perfected to aid scientists in monitoring possible upcoming eruptions \cite{muTomo}. Our use for the Volcanoes is to provide energy discrimination for muon events, thereby aiding in Cosmic Ray studies. HAWC has two nearby volcanoes, Pico de Orizaba and Sierra Negra that could be studied. Both of these volcanoes occupy a significant portion of HAWC's field of view and are sufficiently thick, requiring near horizontal muons to have a minimum momentum of several TeV to penetrate the smaller of the two. Experiments have been proposed to look for neutrino conversion in mountains by detecting high energy particles produced by decaying tau particles emanating from mountains \cite{TaoNeu}. Nearly horizontal muons we observe from the volcanoes would be a background when searching such neutrino events. The neutrino flux is from such conversions in mountains estimated to be very small and thus beyond the scope of this paper, however is of interest to HAWC \cite{neuHAWC}.
 
The focus of this proceeding paper is to present an early analysis of the near horizontal muon events observed by HAWC. A measurement of the single muon rate from different azimuthal angles will be reported. The nearby volcanoes provide varying material overburden depths as a function of arrival direction. Therefore measuring the muon trajectory with sufficient precision in azimuth and zenith enables a measurement of muon flux as a function of overburden depth. Overburden depth serves as a proxy for minimum muon energy thereby enabling a measurement of the muon energy spectrum. The current analysis is done on a limited amount of data nor has it achieved sufficient precision in reconstructing arrival direction to report on a measurement of the muon energy spectrum. Composition studies utilizing muon multiplicity measurements could also be developed as an extension to this analysis in the future. Several multiple muon events have been observed in the data set. An example of such a multiple muon event is shown in a later section.

\section{HAWC}

The HAWC Observatory site and detector is described in detail in another ICRC2017 report \cite{MuonSim}. The completed HAWC detector has been continuously taking data since March of 2015, providing over two years of archived data. However, this uses a data set containing events from 01-01-2016 to 20-01-2016. HAWC is comprised of 300 water tanks, each 7.3 m in diameter and 4.5 m tall, densely packed over a physical area of 22,000 m$^{2}$. Each water tank contains four Hamamatsu Photo-Multiplier tubes (PMTs): one 10 inch at the center and three 8 inch placed around the center evenly. The photomultiplier tubes are capable of resolving time of arrival of Cherenkov light pulses to better than a nanosecond. This time resolution of the PMTs enables the tracking of charged particles as they traverse the tanks of HAWC. The water tanks are aligned roughly in columns and rows slightly rotated by about 10 degrees West of North. The tank elevations are roughly constant. This configuration of isolated tanks allows the detector to be operated as a muon hodoscope that can identify muons traversing from tank to tank at a particle propagation speed consistent with the speed of light. Enhancements of response to certain azimuth directions arise due to the alignment of the tanks.

The available archived data we have to analyze for near horizontal muon events comes from the normal operations. This archived data is optimized to select Gamma Ray and Cosmic Ray extensive air showers for the primary science goals of HAWC. This, unfortunately, is an inefficient trigger for near horizontal muon event detection. The EAS optimized trigger will preferentially select muons with large zenith angles, greater than 85 degrees, which are longer tracks. So the rate of observed muons will be lower than Monte Carlo predictions, and it will be difficult to estimate the true acceptance for these events. Work is now being done to create an additional separate muon optimized event trigger.

The remnant particles from the EAS core are not expected to arrive with the near horizontal muon due to the atmospheric depth. The atmospheric depth for near horizontal muons at HAWC, 4,100 a.s.l., ranges from approximately 3,500 g/cm$^{2}$ at 80 degrees zenith to 22,000 g/cm$^{2}$ at 90 degrees compared to 6,000 to 36,000 g/cm$^{2}$ at sea level respectively. Due to the atmospheric depth at the large zenith angles under consideration in this analysis, we expect to observe only muons from the cosmic ray air shower at the HAWC observation level. 

\begin{figure}[h]
	\centering
	\includegraphics[width=0.75\linewidth]{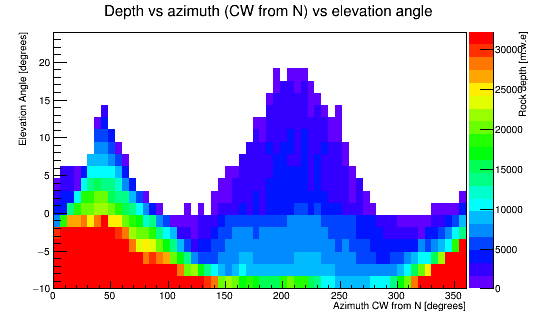}
	\caption{The rock overburden for Pico de Orizaba is centered at 45 degrees azimuth clockwise from north, and the larger of the two, Sierra Negra is centered at 215 degrees azimuth. Pico de Orizaba is about 4 times as distant from HAWC as Sierra Negra.}
	\label{RockDep}
\end{figure}

In the standard science topic analyses, HAWC does not need to worry about two prominent geographical features: the Pico de Orizaba and Sierra Negra volcanoes. This is because both volcanoes, with a maximum height of 20 degrees above the horizon, do not infringe on the nominal air shower events that HAWC observes. This analysis, focusing on near horizontal muons, must account for the volcanoes and the rock depth they will provide. This makes HAWC unique to previous experiments; we are simultaneously both on the surface while viewing significant depths (kilometers) of material overburden. The viewed rock depth depends on a chosen arrival direction and varies from zero meters to about 2000 meters for Sierra Negra and up to 12000 meters for Pico de Orizaba. The rock depths with respect to a chosen arrival direction are shown in Figure \ref{RockDep} with data from Mexico's National Institute of Statistics and Geography \cite{RockDepData}.

\section{MUON DETECTION}

\begin{figure}[h]
	\centering
	\includegraphics[width=0.49\linewidth]{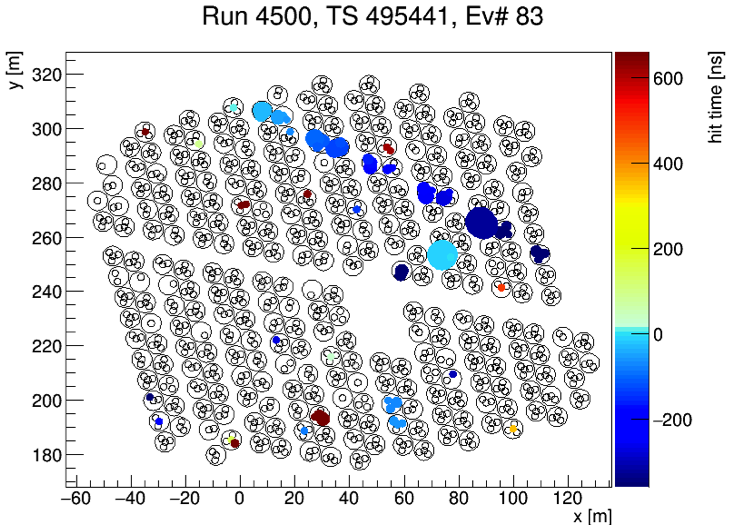}
	\includegraphics[width=0.49\textwidth]{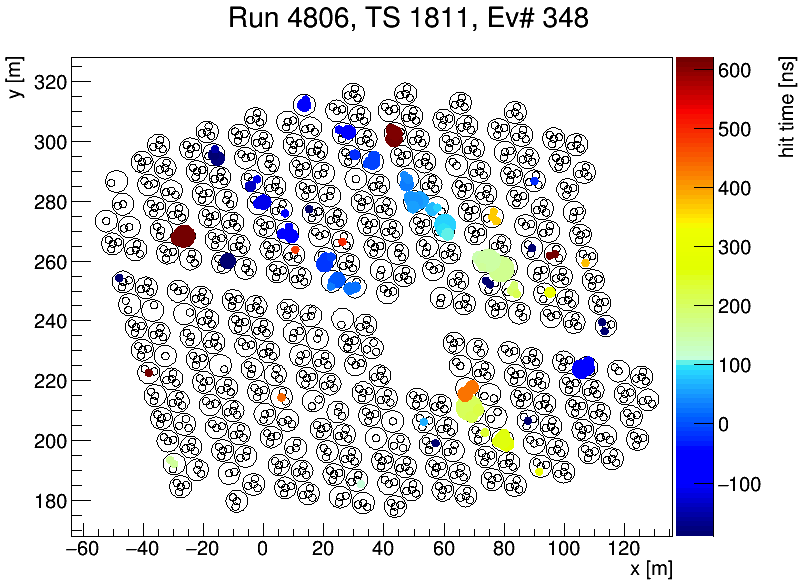}
	\caption{(left) This is an event with a single muon crossing the northeast portion of HAWC with a zenith angle greater than 88 degrees. (right) This is a double muon event traversing the breadth of HAWC starting from the northwest. This double muon event was discovered when performing a visual inspection to identify any background events.}
	\label{muon-exam}
\end{figure}

To look for near horizontal muons, like those seen in Figure \ref{muon-exam}, HAWC will be treated as a hodoscope by using the closely packed tanks to detect single or multiple particles traveling through neighboring volumes of water. However, the layout of the tanks is in triangularly packed rows two tanks wide. This hodoscope detector will allow for an enhanced detection for some trajectories. In all there are 3 orientations to have an enhancement due to the triangular geometry, in all six azimuthal trajectories. From Monte Carlo simulations of single muons, using the HAWC array, these six trajectories are statistically equal (in pairs) with each other if there are no obstructions. The differences between the three orientations are due to the cross-sectional area. We do not expect the muon rates for each of the directions to be equal due to a different amount of overburden and the earth's magnetic field influence.

The data sets we have access to are composed almost entirely of extensive air shower events. Each triggered event requires that there are at least 24 PMTs within a 150 nanosecond time window, which HAWC has optimized for the primary science goals. Muon events which are greater than 50 meters\footnote{The requirement for 50 meters is the distance a high zenith angle muon (>84.5 degrees) is able to travel in the requisite 150 ns time window.} may pass this cut due to most azimuth angles having about 24 PMTs in that distance; PMT shot noise and single tank vertical muons assist in the triggering of these events. The first cut placed on events when we later process the data for muons is requiring fewer than 50 PMTs flagged in the trigger. That cut quickly eliminates the larger EAS background events.

The first step to identify an event as a muon uses the time and position information from PMT hits to find those which are consistent with the particle propagation at the speed of light with respect to a reference PMT. This step is independant of any propagation direction and thus finds many EAS events that have ``speed of light like'' hits. This is not because the event is actually traveling at the speed of light with respect to the reference PMT; there are enough random hits that satisfy a speed of light cut. These background events have a subset of the PMTs in the EAS selected, which then form arcs and circles typically centered on the reference PMT and are eliminated by using a Hough Transform. 

The Hough Transform is a technique used in image processing to identify features like lines, the simplest feature, to more complicated shapes like circles and squares \cite{houghtransf}. We transform the plot of the hit PMT locations in a Cartesian coordinate space (Y vs X) into a  plot using  a coordinate system of R, distance of closest approach to the center of HAWC along a line drawn in a given direction, $\theta$. Near horizontal muons will trigger PMTs along a narrow line without bending, as the high energy muon is not likely to scatter in another direction. We require that a muon event have at least 70\% of the PMTs align within one ( $\theta$, R ) bin 1 degree and 5 meters. Small EAS events masquerading as a muon will have a wider distance distribution and fail the Hough Transform. 

\section{RESULTS}

The near horizontal muon rate HAWC observes, independent of direction, is calculated every 10 minutes. The rate is 1.25 Hz with a standard deviation of 0.09 resulting in an approximate 108,000 muons per day. This observed rate is using strict cuts and affected by the inefficiency due to the nominal trigger. The current cuts preferentially choose muons with zenith angles greater than 85 degrees, which are longer muon tracks in HAWC. Currently, it is difficult to say what the efficiency for muon detection is due to simulations being in progress and highly dependent on the azimuth and zenith for a muon event \cite{MuonSim}. 

\begin{figure}
	\centering
	\includegraphics[width=0.7\textwidth]{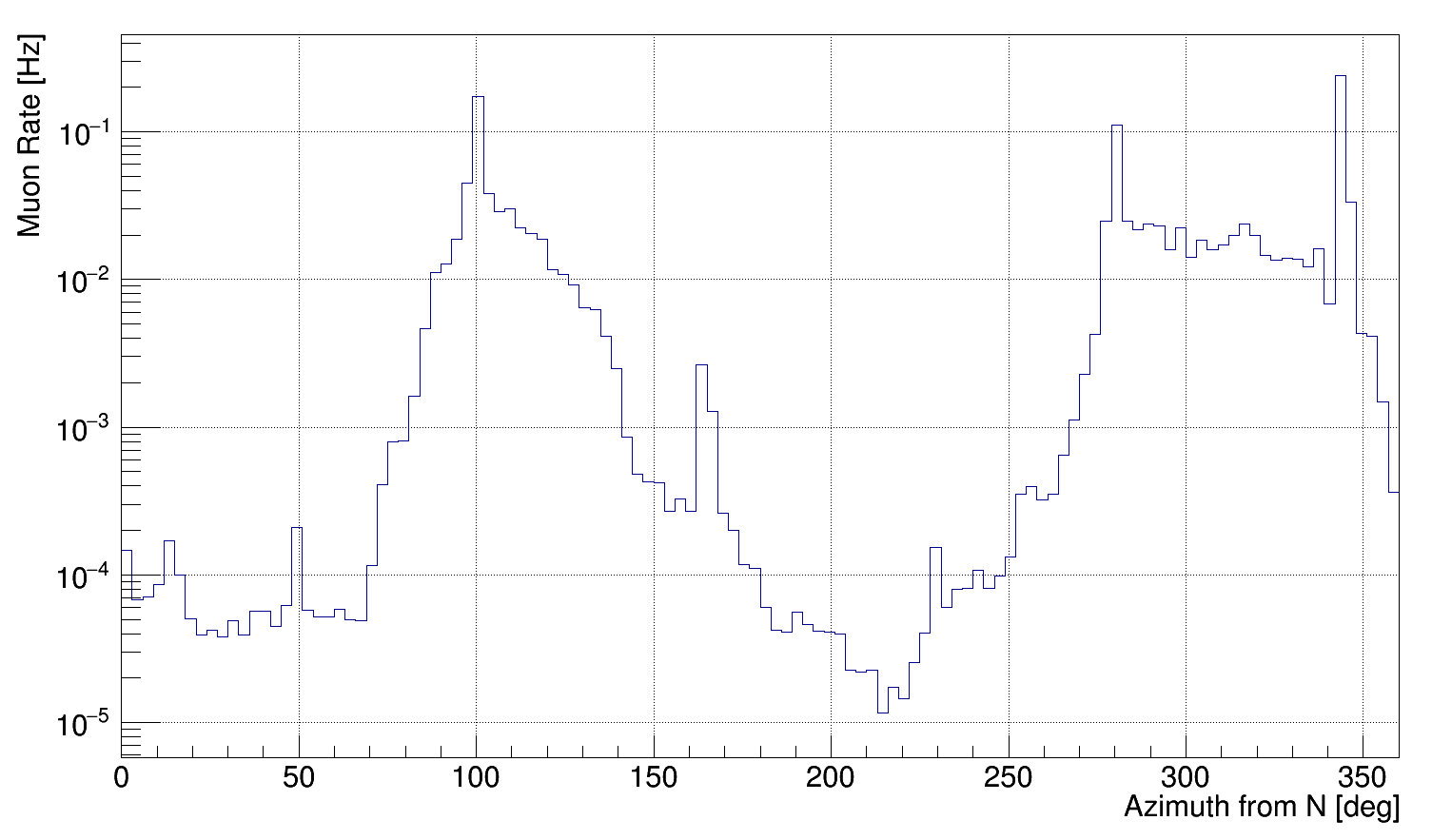}
	\caption{The observed muon rate for each azimuth direction and zenith angles above 85 degrees.}
	\label{azimuth}
\end{figure}

The observed near horizontal single muon rate as a function of azimuthal direction for the processed data is shown in Figure \ref{azimuth}.  One feature to note is the increased rate in certain preferred directions due to the instrumental response enhancement for directions where HAWC tanks are aligned. These preferred directions and rates are indicated in Table \ref{EasyDirections}.  For instance, the expected rates from the North and South enhancement directions (345 degrees and 165 degrees azimuth) are different due to the difference in rock overburden.  Muons traveling from  the north direction  are unimpeded by rock overburden, which is not true for those coming from the south direction. The rate from the peak at 165 degrees azimuth (the south) is fairly high for what we expect with the rock depth of approximately 2500 m.w.e in the direction near 90 degrees zenith. Our analysis detects muons with a minimum track length of 50 meters; this corresponds to a probable zenith very near 85 degrees. We observe a second feature in the plot of  rate as a function of arrival direction.  The effect of the overburden depth in the direction of the volcanoes results in a reduction in the rate of muons. The reduction in rate roughly matches what would be expected from the overburden depth, Figure \ref{RockDep}, as a function of arrival direction.  The comparison of the rates from directions with significantly different overburden should allow HAWC to measure the integrated muon flux above several different energies. Thereby allowing us to estimate the primary cosmic ray energy spectrum from observed muon rates using detailed simulations of the instrumental response of HAWC and the effect of overburden.

\begin{table}[h]
	\centering
    \begin{tabular}{| l | l | l | l |}
    \hline
    Azimuth & Approximate direction & Rate & Rock depth at 90 degrees \\ \hline
    47 & Northeast to Southwest & 0.21 mHz & 7000 m \\
    227 & Southwest to Northeast & 0.15 mHz & 2500 m \\ \hline
    165 & South to North & 2.7 mHz & 1500 m  \\ 
    345 & North to South & 0.24 Hz & 0 m \\ \hline
    105 & East to West & 0.17 Hz & 0 m \\
    285 & West to East & 0.11 Hz & 0 m \\
    \hline
    \end{tabular}
    \caption{These are the six preferred directions where the HAWC tanks easily align and have a detection enhancement. } 
    \label{EasyDirections}
\end{table}

\section{OUTLOOK}

Additional work for this near horizontal muon analysis is to analyze the raw HAWC data before the nominal event trigger. This to eliminate the inefficiencies of finding muon events from EAS optimized trigger data. Many, if not most, of the near horizontal muons are being eliminated from what we have learned when comparing to the toy model simulation. The new trigger will be able to find events with zenith angles near 80 degrees, thereby making HAWC's cross sectional area larger and increasing the rate HAWC will observe shorter track muons and muon bundles. By having shorter tracks, the bias will be to reconstruct events on the preferred azimuthal directions. This is because of the HAWC tanks and the several meter wide aisles between tank rows; the angular resolution depends strongly on the track length and will require PMT charges to more precisely define the trajectory through the tank.

One of the next steps is to improve the multiple muon identification algorithm. Currently the Hough Transform is able to suggest if there are multiple tracks, although, it is currently only able to find the best muon. The Hough Transform is able to identify multiple lines automatically through the binning threshold, but those are only possible muon tracks and we must process them as candidates. For instance: two parallel lines of the same length would appear at the same angle and with the same bin count; but would have different distance offsets. Multiple muon events observed in simulated data show clearly that while the tracks may be parallel, they would likely contain different numbers of PMTs because each muon will pass through a slightly different region of HAWC. 

We are looking to turn these muon detection rates into a possible estimate of the cosmic ray flux. The volcanoes will provide a minimum threshold for muon energy. The muon rate for this threshold is correlated to the integrated cosmic ray energy spectrum. In addition, multiple muon events appear to be fairly common in our data set. We are confident that we will have a sufficient rate for future analyses. The muon bundles are also a possible characteristic of different primary particle species. The study of cosmic ray composition will require a careful observation of how muon bundles vary with increasing primary particle energy. 


\section*{Acknowledgments}
\footnotesize{
We acknowledge the support from: the US National Science Foundation (NSF); the US Department of Energy Office of High-Energy Physics; the Laboratory Directed Research and Development (LDRD) program of Los Alamos National Laboratory; Consejo Nacional de Ciencia y Tecnolog\'{\i}a (CONACyT), M{\'e}xico (grants 271051, 232656, 260378, 179588, 239762, 254964, 271737, 258865, 243290, 132197), Laboratorio Nacional HAWC de rayos gamma; L'OREAL Fellowship for Women in Science 2014; Red HAWC, M{\'e}xico; DGAPA-UNAM (grants RG100414, IN111315, IN111716-3, IA102715, 109916, IA102917); VIEP-BUAP; PIFI 2012, 2013, PROFOCIE 2014, 2015;the University of Wisconsin Alumni Research Foundation; the Institute of Geophysics, Planetary Physics, and Signatures at Los Alamos National Laboratory; Polish Science Centre grant DEC-2014/13/B/ST9/945; Coordinaci{\'o}n de la Investigaci{\'o}n Cient\'{\i}fica de la Universidad Michoacana. Thanks to Luciano D\'{\i}az and Eduardo Murrieta for technical support.
}

\end{document}